\documentclass[a4paper,twocolumn,DIV=16]{scrartcl}

\usepackage[normalem]{ulem}

\usepackage{graphicx}
\usepackage{amsmath}
\usepackage{url}
\usepackage{hyperref}
\usepackage{cleveref}
\usepackage{microtype}
\usepackage{abstract}
\usepackage[textsize=footnotesize,obeyDraft,draft]{todonotes}
\usepackage[all]{nowidow}

\begin{document}

\title{A software controlled voltage tuning system using multi-purpose ring
oscillators}

\date{March 2015}
\author{Steve Kerrison and Kerstin Eder\\
    University of Bristol\\
    \texttt{\{steve.kerrison,kerstin.eder\}@bristol.ac.uk}
}
\maketitle

\begin{abstract}

This paper presents a novel software driven voltage tuning method that utilises
multi-purpose Ring Oscillators (ROs) to provide process variation and
environment sensitive energy reductions. The proposed technique enables voltage
tuning based on the observed frequency of the ROs, taken as a representation of
the device speed and used to estimate a safe minimum operating voltage at a
given core frequency. A conservative linear relationship between RO frequency
and silicon speed is used to approximate the critical path of the processor.

Using a multi-purpose RO not specifically implemented for critical path
characterisation is a unique approach to voltage tuning. The parameters
governing the relationship between RO and silicon speed are obtained through the
testing of a sample of processors from different wafer regions. These parameters
can then be used on all devices of that model. The tuning method and software
control framework is demonstrated on a sample of XMOS XS1-U8A-64 embedded
microprocessors, yielding a dynamic power saving of up to 25\% with no
performance reduction and no negative impact on the real-time constraints of the
embedded software running on the processor.
\end{abstract}

\section{Introduction}

Modern embedded computing systems require ever-increasing performance from
microprocessors whilst simultaneously consuming less energy. Progress in both of
these areas leads to new opportunities in embedded applications.

Advances in silicon fabrication technologies bring shrinks in feature size,
which along with increased transistor counts, helps to reduce power by needing a
lower operating voltage than the previous generation of devices. However, the
relationship between feature size, power and performance becomes more complex
and subject to greater variability with smaller process technologies, and so
feature size alone cannot be relied upon to improve both performance and power.

Techniques such as Dynamic Voltage and Frequency Scaling
(DVFS)~\cite{burd2000scaled}, power- and clock-gating, and advanced sleep
states~\cite{pgsleep} are used in combination with operating systems and
application software to ensure energy consumption is minimised whilst delivering
sufficient performance for a set of tasks.

In order for a processor manufacturer to ship a product and guarantee
reliability, it must select operating parameters that will enable the product to
function correctly in spite of any process variation across the range of parts.
As such, the operating voltages in the product data sheet must be chosen with a
degree of conservatism. Additionally, products may be \emph{binned} into
different speed or power categories, depending on their behaviour under test,
but there will still be variation, even across the binned parts, which presents
challenges in categorising parts reliably~\cite{Sartori2010}.

Ring Oscillators (ROs) are free-running logic blocks that operate at a frequency
governed by the delays intrinsic to their structure and also the behaviour of
the silicon within which they are implemented. Therefore, their behaviour varies
from chip to chip. This can be exploited in various ways, for example as a
random number seed, or for clock generation. In this paper, ROs combined with
hardware counters are used to determine the most appropriate operating voltage
for a processor given the observed RO speed and the desired operating frequency.

The XMOS XS1-U8A-64 processor is used as the test subject for this technique, by
virtue of its voltage scaling capabilities and RO implementation. The ROs are
embedded in the processor silicon, but are separate components within the
processor accessible via software and so can be considered multi-purpose.

This paper makes the following contributions to the areas of energy efficient
embedded software/hardware co-design and embedded processor architectures:

\begin{itemize}
\item A unique application of multi-purpose software-controlled ROs, rather than
custom hardware blocks.
\item A flexible soft control loop for voltage tuning a system, that is both
process variation and temperature sensitive.
\item The control method, although implemented in software, has zero impact on
the timing of an application running on the processor.
\item Granularity of control is unconstrained in software as characterisation
formulae are used rather than a table of operating states.
\item The method's power saving capabilities are demonstrated, through testing
and evaluation on a set of samples of the target device.
\end{itemize}

The rest of this paper is organised as follows.  Section~\ref{sec:bg} explores
frequency and voltage scaling, existing voltage tuning approaches including
methods for evaluating the silicon speed, and uses of ROs in processors. In
Section~\ref{sec:imp} a new tuning approach is described in the context of the
chosen target hardware. Section~\ref{sec:test} shows the results of testing this
technique on a sample of target processors.  An evaluation of the
technique's effectiveness is presented in Section~\ref{sec:ev}, followed by
discussion of future work in Section~\ref{sec:future} and conclusions in
Section~\ref{sec:conclusions}.

For clarity, this paper refers to \emph{energy} and \emph{power} in terms of
\emph{consumption} and \emph{dissipation} respectively. That is, \emph{energy
consumption} is a measure of \emph{total work done}~---~the amount of
potential that is transformed in order to achieve the desired outcome and
typically measured in Joules. \emph{Power dissipation} is an instantaneous
measure of a rate of energy transfer, expressed in Watts. Power dissipation at
1 Watt for 1 second results in an energy consumption of 1 Joule. The majority of
this paper refers to power, rather than energy, for consistency.

\section{Background}
\label{sec:bg}

This paper builds upon research into and application of techniques in the areas
of CMOS device properties, voltage and frequency scaling, and ROs.
This section covers the relevant background within these three areas.

\subsection{Power dissipation and DVFS}

The technique of Dynamic Voltage and Frequency Scaling (DVFS) is motivated by a
desire to minimise energy consumption in a device by operating in the most
efficient possible trade-off of power vs. performance for a given
workload~\cite{burd2000scaled}. DVFS is affected mainly by two components of
power dissipation in a CMOS device: static and dynamic power.

\subsubsection*{Static power}

The main component of static power is the leakage current of the transistors in
the silicon. This is present regardless of the on/off state of transistors. As
processors are fabricated on smaller process nodes, the percentage of overall
power dissipation that is attributed to leakage is
growing~\cite{kim2003leakage}, for example due to increased leakage through
thinner gate oxide layers, which must be combated with technology such as
improved high-k gate dielectrics~\cite{highkreview}.

\begin{equation}
P_s = VI_{\text{leak}}
\label{eq:static}
\end{equation}

In Equation~\ref{eq:static}, the static power, $P_s$, is the product of the
device voltage, $V$, and the leakage current, $I_{\text{leak}}$.  Thus, there is
a simple linear relationship between operating voltage and static power.

\subsubsection*{Dynamic power}

Power dissipated in order to switch transistors on or off is termed dynamic
power, $P_d$, and is expressed in Equation~\ref{eq:dynamic}.

\begin{equation}
P_d = \alpha C_\text{sw}V^{2}F
\label{eq:dynamic}
\end{equation}

$C_\text{sw}$ is the capacitance of the transistors in the device and $\alpha$
is an activity factor or the proportion of them that are switched. Activity
factor is workload specific, but often estimated as switching half of the
transistors in the device~\cite{Brooks2000}, giving $\alpha = 0.5$. $F$ is the
operating frequency of the device. Observe that changes in $V$ have the biggest
influence on dynamic power dissipation.

A reduction in $V$, however, will slow the transistor switching speed,
increasing the delay in the critical path, requiring that $F$ also be lowered.
Thus, there is a trade-off between reduced power dissipation and the total
energy consumption due to longer execution time~---~in some cases it is not
beneficial to slow the device down further. Choosing a strategy for energy
saving, be it tuning the frequency to avoid slack time, or racing to idle by
operating at high speed briefly, then reducing to a low power state, is
dependent on the type of work and the behaviour of the system; there is not one
strategy that works in all cases~\cite{Amur2008}.

The relationship between voltage and frequency varies depending on manufacturing
process and device implementation. Simplistic representations, such as that
in~\cite{kim2003leakage}, represent the relationship as $F \propto \frac{V -
V_{\text{th}}}{V}$, where $V_{\text{th}}$ is the threshold voltage of the
transistor. As $V$ approaches $V_\text{th}$, $F$ approaches zero. The nominal
operating frequency and voltage, $F_{\text{norm}}$ and $V_{\text{norm}}$
respectively, can therefore be represented as Equation~\ref{eq:norms}, taken
from~\cite{kim2003leakage}, where $V_{\text{max}}$ is the maximum operating
voltage of the transistor.

\begin{equation}
V_{\text{norm}} = 
    F_{\text{norm}}\left(1 - \frac{V_{\text{th}}}{V_\text{max}}\right)
    + \frac{V_{\text{th}}}{V_{\text{max}}}
\label{eq:norms}
\end{equation}

A step reduction in frequency yields a smaller step reduction in voltage. With a
conservative view, where preserving correct operation is required, the
relationship can be represented linearly.

\subsubsection*{Other losses}

Conditions such as short-circuit current can also be factored into the overall
power dissipation of a device. Techniques such as the $\alpha$-power law MOS
model consider these~\cite{Sakurai1988}. In this paper, however, these
additional effects are considered to be part of either dynamic or static power,
depending on their relationship to transistor switching activity.

\subsection{Ring Oscillators}

An RO is typically implemented as a series of connected inverters, with the
final inverter's output looped back to the input of the first. Provided an
odd number of inverters are used, the circuit will be astable and the output
will switch states continuously at a frequency governed by the propagation
delays in the inverters and their connecting wires.

The simplest model for the frequency of an oscillator is determined by the
number of inverters that form it~\cite{Mandal2010}. Equation~\ref{eq:rospeed}
expresses a RO's frequency, $F_{\text{o}}$, as the number of inverters,
$N$, and the propagation delay of each inverter, $D_{\text{inv}}$, where $2$
inversions produce a cycle.

\begin{equation}
F_{\text{o}} = \frac{1}{N \times 2D_{\text{inv}}}
\label{eq:rospeed}
\end{equation}

The delay term, $D_{\text{inv}}$, is dependent on multiple factors, including
manufacturing process, transistor size, operating voltage and device
temperature. This makes a RO unsuitable on its own as a stable clock source,
but creates various other possible applications.

\begin{figure}[]
\centering
\includegraphics[trim=0cm 0cm 0cm 0cm,clip=true,width=3in]{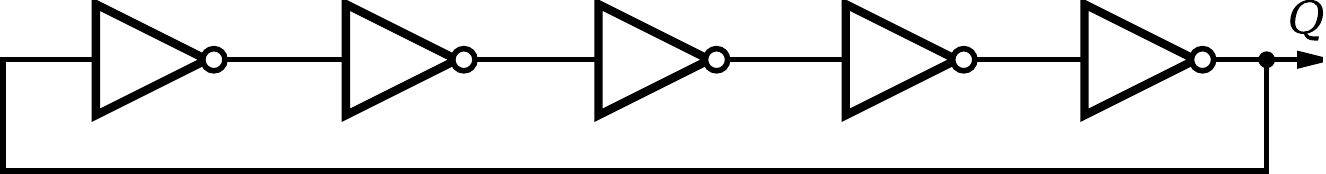}
\caption{Basic construction of a ring oscillator, with feedback of output
\emph{Q} to the first inverter in the chain. Wire lengths and the number of
inverters affect the oscillator frequency.}
\label{fig:RO}
\end{figure}

ROs can be used for a wide range of purposes, including as an entropy source for
hardware random number generation~\cite{XMOS2010a}, as voltage controlled
oscillators within PLLs (Phase Locked Loops)~\cite{Weigandt1994}, or as part of
control circuitry for voltage-sensitive components~\cite{burd2000scaled}.

\subsection{Frequency and voltage selection, critical path estimation}

In a typical use of DVFS in general purpose computing, the operating system will
instruct a processor to change its power state depending on its
workload~\cite{linuxgov2013}. This will trigger a change in frequency and/or
voltage, balancing performance with energy consumption. The voltage and
frequency points are typically selected by the processor manufacturer and must
ensure valid operation for all processors of a given model. As such, they must
be sufficiently conservative to account for process variation from manufacturing
the processors, where the location of the die on the wafer may affect its speed.
The challenge lies in monitoring or correctly modelling the critical path or
paths in the processor. In embedded and deeply embedded systems, DVFS may be
applied using different constraints, or without the assistance of an operating
system, but shares the same power-saving goals.

Various hardware-assisted approaches for voltage tuning exist. In-situ error
detectors can be placed into a processor design~\cite{das2006dvstuning}. These
detectors can identify when the voltage is too low (or the frequency is too
high), and then appropriate action can be taken to correct the timing issue and
re-execute any failed instructions.

A delay line can be used to characterise the critical path.
In~\cite{Ikenaga2011}, a \emph{Universal Delay Line} (UDL) is introduced, which
aims to be portable across designs by containing a gate structure that minimises
delay error and thus act as a reliable input for voltage control. Multiple UDLs
are used to account for within-die variation. The reported results demonstrate
that voltage tuning using this method achieves a 27\% active power reduction.

A similar approach has been used in Field Programmable Gate Arrays (FPGAs), in
which a delay line was implemented that was timed in order to establish whether
the FPGA fabric was operating quickly enough, or needed additional
voltage~\cite{Nabina2012}. This allows the FPGA to provide a reconfigurable
hardware control module to a system, with tightly tuned voltage and frequency
scaling capabilities. Further research embeds in-situ detectors into arbitrary
IP blocks targeting an FPGA, to achieve a similar goal to the delay line
approach, but more closely integrated with the target IP~\cite{Nunez-Yanez2013}.
The DVFS control is tightly connected with the in-situ error detection circuitry
to ensure that changing operating conditions do not lead to unrecoverable
errors.

Critical paths can be estimated via other methods, such as~\cite{Liu2010}, in
which the multiple possible critical paths of a complex processor, combined with
the variations introduced from modern silicon process technologies, are used to
create a representative model. This reflects the worst case delay of the circuit
and is shown to have an average error margin of less than 2.8\%, with a lower
level of pessimism required than other estimation methods of capturing the
critical path delay. 

Ring Oscillators can be used as part of a control loop in a DVFS
implementation~\cite{burd2000scaled}. The RO can be used for directing a voltage
controller when frequency changes are requested. Changes in RO speed are taken
to be a simplified analogue of changes in the critical path of the hardware,
forming part of the feedback loop to the frequency selector, which adjusts the
supply voltage until the target frequency is reached.

Other approaches, such as that of \cite{lee2000hopping}, instead implement a
selection of frequencies and voltage in a table, controlled by software, which
can schedule changes to the frequency and voltage based on the worst-case
execution time of a set of tasks that form a workload.

In power-gated circuits, the gate sizing can be exploited as a method of
adaptive power control. In~\cite{Hsieh2011}, a network of power gates are
selectively enabled. A smaller number of enabled gates limits the voltage
supplied to the connected logic. Device activity is monitored by measuring
supply voltage, where a period of switching activity will result in a dip in
voltage, followed by a return to previous levels, thus the loading of the
circuit can be observed and the voltage during slack periods optimised,
resulting in a 12\% average power reduction.

\subsubsection*{Comparison}

The key differences between our contribution and prior work are, firstly, that
the proposed implementation utilises an existing hardware block that is designed
for multiple functions, not specifically as a critical path model or as part of
a hardware control loop. Secondly the proposed voltage tuning approach forms a
hardware/software control loop, in which the voltage selection decisions, as
well as safety margins, are implemented in software. Further, the control
algorithm uses characterisation formulae, rather than table look-ups, to provide
a target voltage, thus imposes no software restriction on the number of possible
voltage/frequency selections.

These differences provide greater flexibility and potential portability to other
systems than related work. However, the latency increase incurred from
implementing the control system in software limits the ability to save power
over fine-grained time intervals, and the simpler hardware block used to
represent the critical path necessitates a more conservative safety margin.
Possible improvements to these areas are discussed in Section~\ref{sec:future}.

\section{Implementation}
\label{sec:imp}

This section describes the selected processor family for use in experimentation,
along with the software technique used to apply RO-based voltage tuning to the
devices.

The following requirements are key to the ability to apply the proposed voltage
tuning technique:

\begin{itemize}
\item A configurable power supply, with sufficiently fine-grained control to
allow changes to the device's core supply without necessarily needing to change
frequency.
\item Configurable frequency, at run or boot time, and ideally dynamically.
\item Internal ROs, attached to hardware counters, to provide assessment of the
device's speed.
\end{itemize}

The XMOS XS1-U8A-64 processor was selected based on these criteria. Other
processors, particularly soft-cores for FPGAs, could also be used, with some
modification to include ROs that can be sampled, using similar methods to the
delay line or in-situ detectors described in~\cite{Nabina2012}
and~\cite{Nunez-Yanez2013}.  However, the XMOS processor has all the required
capabilities readily available.

\subsection{Test device: XMOS XS1-U8A-64 processor}

The XS1-U8A-64 combines a hardware multi-threaded XS1 processor with a set of
peripherals that provide a USB PHY, various analogue components such as ADCs,
and configurable power supplies. 

The XS1 multi-threaded architecture has I/O and peripheral component control
built into the instruction set, rather than memory-mapped. The architecture is
described in more detail in~\cite{XMOS2009a} and~\cite{XS1Lsys2008}. It is used
to implement flexible hardware interfaces in software using a C-like language,
with very low latency (as little as 10~nS) between pin activity and software
response. The predictable timing of the architecture makes it well suited to
hard real time embedded software. Of particular interest to this experiment,
each core has four ROs within it, with two distinct implementations and two
locations in the design.  One of each RO implementation is placed near the I/O
ports of the device, and the other two are located near the processor core.

These ROs act as clock sources for a set of 16-bit hardware counters, which can
be selectively enabled/disabled and the counter values read with a simple
sequence of instructions~\cite{XMOS2010a}. Thus, by enabling a RO's counter for
a specified period, the speed of the RO can be determined. Even without detailed
knowledge of the RO implementation, its speed can be compared to other chips of
the same series, assuming a consistent reference clock for timing.

Peripheral components of the XS1-U8A-64 are presented as endpoints an XMOS
device network, accessible via the channel communications paradigms established
in the XS1 instruction set architecture~\cite{XMOS2009a}. They are configurable
in a similar way to I2C or SPI peripherals, but at the physical level and
low-level in software, the interface is somewhat different.

Three power supplies are provided in the peripheral part of the XS1-U8A-64, one
3.3~V for I/O logic and two 1~V, for separate Phase-Locked Loop (PLL) and core
supplies. For this research, the core supply is the only one that is adjusted.
This particular supply can be configured between 0.6~V and 1.3~V in 10~mV steps,
with a recommended slew rate of 10~mV per microsecond to limit over- and
under-shoot.

Assuming a safe set of default conditions for both the power supply and core
frequency, both can be changed dynamically during program execution. The voltage
should be changed no faster than the aforementioned slew rate, whilst the core
frequency can either be divided to a lower frequency on the fly, or the PLL can
be reprogrammed to a new target frequency~\cite{XS1Lsys2008}, triggering a
soft-reset and reboot of the core.

\subsection{Software requirements and hardware considerations}

A software implementation of self tuning voltages and frequencies must consider
the behaviour and capabilities of the underlying hardware whilst giving certain
assurances to the application software that will be running upon it. An embedded
environment with hard real-time constraints is considered. As such, a number of
criteria must be given attention.

\subsubsection*{Environment and workload affect silicon speed}

Transistor switching speed increases in an approximately linear relationship to
voltage whereas rising temperature can either increase or decrease speed,
depending on the feature size~\cite{cmosTempInversion}. However, higher voltages
result in greater dynamic and static power dissipation, and so the relationships
between design thresholds, workload, speed, voltage and temperature are not
always straightforward. For example, the relationship between temperature and
threshold voltage can typically be represented linearly, but the static current
leakage has an exponential relationship with temperature~\cite{Wolpert2012}.

Processor temperature may be influenced by the ambient temperature of the
operating environment, but also by the workload run upon it, as this will
increase energy consumption and thus power dissipated as heat.

In order to provide a reasonable expectation of safety in a voltage tuned chip,
its speed should either be constantly monitored, or if this is not possible, it
should be measured an appropriate limit of its operating temperature in the
given environment. In the latter case, an environmental change may lead to a
fault or sub-optimal energy usage.

\subsubsection*{Performance cannot be impacted by voltage tuning}

If a given application is analysed and proven to work at a particular operating
frequency, then the introduction of voltage tuning should not adversely affect
that. This constrains the tuning to finding the lowest voltage for the currently
assigned frequency. Other strategies may be acceptable in other workloads, such
as finding a suitable frequency for a given voltage, in an environment where the
nominal voltage may not be achievable. However, for this paper, the focus is
upon tuning the voltage to the current frequency.

\subsubsection*{Latency and deadlines cannot be adversely affected}

A fine-grained performance requirement in an embedded system is that response
times to certain events must be kept low in order for hard-deadlines to be met.
As such, the process of monitoring the silicon speed or changing the voltage
must not cause deadlines to be missed. The simplest method for guaranteeing this
is to avoid any activity that would affect timing in any way, such as inserting
additional tasks into the workload, or modifying existing tasks.

\subsection{Selected approach}

Based on the discussed criteria, the following implementation details are used
in the voltage tuning framework:

\subsubsection*{Silicon speed will be profiled and a new voltage applied before
main application execution.}

This avoids any performance or fine-grained timing issues by not introducing any
extra processing during execution of the main application. The analysis time and
power supply slew rate become decoupled from the constraints of the program.
However the effect upon start-up time may be undesirable for applications
requiring a very rapid cold-start. It may also fail to account for environmental
changes, such as dramatic ambient temperature variation.

\subsubsection*{A self-exercising routine will heat the processor before
tuning.}

In order to ensure the speed of the device is measured appropriately, a
high-power test loop will be executed for a period of time before and continue
throughout the speed profiling phase. This heats up the silicon prior to testing
and keeps it warm during. This approach assumes that the chip package and
circuit board's heat dissipation, as well as the environment and processor
workload do not make for a gradual, unabated rise in operating temperature over
a longer time period.

\begin{figure}[htbp]
\centerline{\includegraphics[trim=1cm 11cm 0.5cm
1cm,clip=true,width=3.5in]{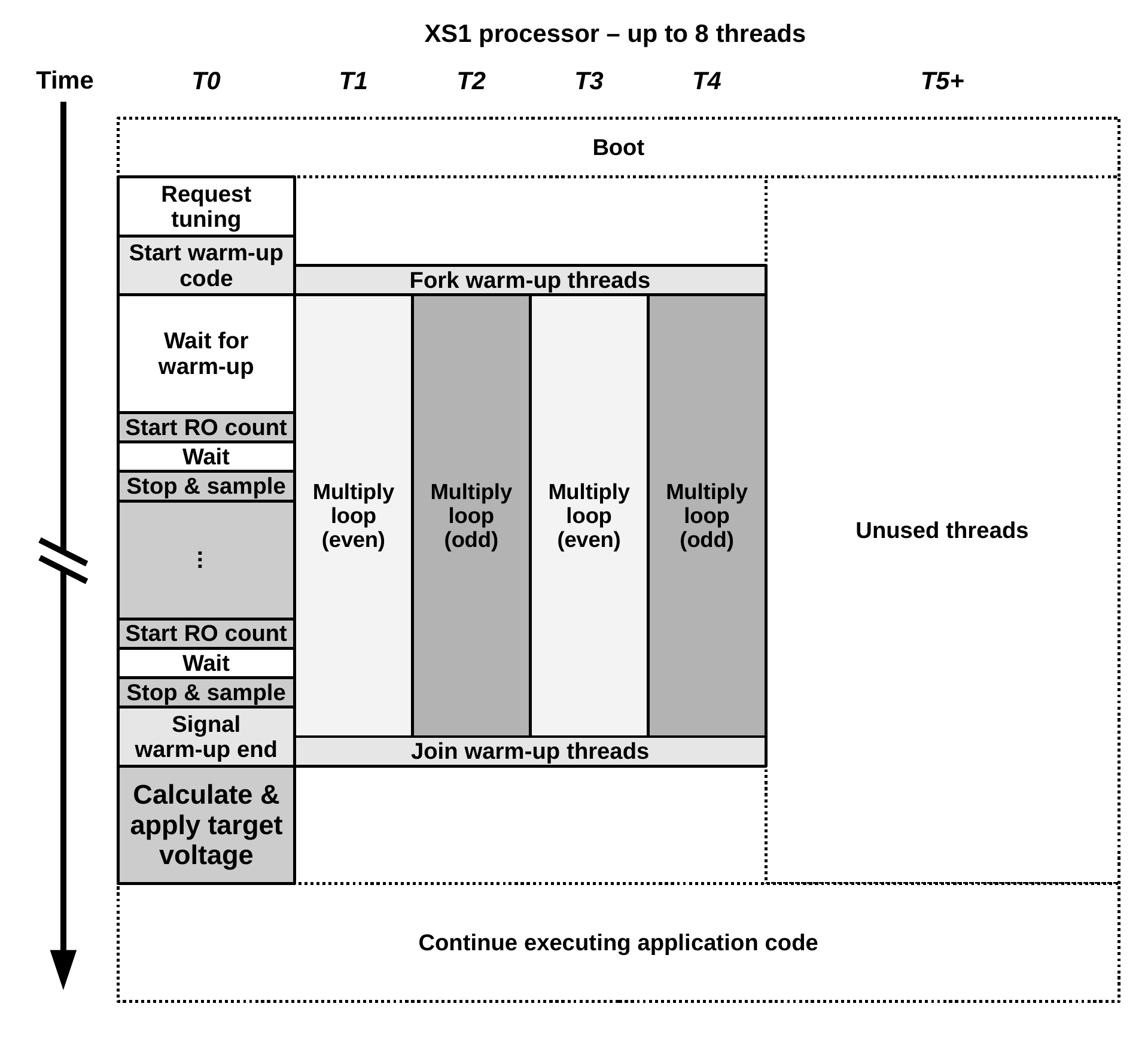}}
\caption{Depiction of voltage tuning process, including warm-up and RO sampling,
followed by normal application execution.}
\label{fig:framework}
\end{figure}

The framework is implemented as a series of libraries that provide control over
the core voltage, routines for heating up the processor and a process for
measuring the silicon speed and selecting an appropriate voltage for the given
frequency. The implementation is outlined in Figure~\ref{fig:framework} and
an explanation follows.

The simplest invocation of the framework is to request that it set the core
voltage to the lowest safe level for the given clock speed. When doing this, the
framework first determines the frequency by reading the PLL configuration and
core clock divider in combination with a compile time macro that specifies the
oscillator frequency. This assumes, therefore, that the oscillator frequency is
correctly specified by the hardware designer and/or software developer.

Next, the processor is heated for a period of time which at its 65~nm feature
size, will slow the ROs~\cite{cmosTempInversion}. This aims to reflect the
silicon speed under a heavy workload. This is achieved by executing several
threads of interleaved multiplication operations with specially selected operand
values. This particular set of threads has been shown to maximise the power
dissipation of the core~\cite{Kerrison2013}.

After a warm-up period of one second, the values of the counters connected to
the ROs are recorded, then the counters enabled, incrementing at a rate governed
by the RO frequency. In testing, one second was found to be sufficient heating
time to produce the observable RO slowdown. After 85~$\mu$S, the counters are
stopped, re-read and the difference calculated. This measurement duration
captures a good sample of the RO frequency, without overrunning the 16-bit
counters.  This measurement step can be performed several times to establish an
average.  The thread responsible for configuration and sampling of the ROs is
inactive the majority of the time, leaving the warm-up threads to fully occupy
the pipeline.  Once sufficient samples are collected, the slowest of the
device's ROs is then taken as an approximation of the silicon speed.

Two RO scaling characteristics, $S_f$ and $S_v$, must be determined, with the
aim of providing an analog for Equation~\ref{eq:norms}, to give a target
operating voltage and/or frequency (equivalent to determining appropriate
$F_\text{norm}$ and $V_\text{norm}$). $S_f$ is the ratio between RO frequency,
$F_o$, and processor frequency, $F_p$, such that $F_o$ indicates that the
silicon is operating quickly enough to meet the timing requirements of the
processor at $F_p$. This yields the inequality in Equation~\ref{eq:tgtosc},
which states the minimum $F_o$ for a given target processor frequency. The
second characteristic, $S_v$, is the ratio between core voltage and $F_o$,
satisfying Equation~\ref{eq:sv}.

\begin{align}
F_o &\ge \frac{F_p}{S_f}
\label{eq:tgtosc}
\\
V &\ge F_o \cdot S_v
\label{eq:sv}
\end{align}

If $F_o$ is the current RO frequency and $V$ is the current core voltage, then a
new target RO frequency, $F_o^\prime$, may be found that still satisfies
Equation~\ref{eq:tgtosc} and similarly a new voltage, $V^\prime$, that can
provide $F_o^\prime$, per Equation~\ref{eq:sv}.

In the above example it is assumed that the processor is operating safely and
that a voltage \emph{optimisation} is taking place. It is also possible to
calculate a higher $V^\prime$ for a higher target processor frequency,
$F_p^\prime$, using the same method. In either case, $V^\prime$ is calculated
using Equation~\ref{eq:vtgt}.

\begin{equation}
V^\prime = V + S_v \cdot \left(F_o^\prime - F_o\right)
\label{eq:vtgt}
\end{equation}

The framework's interface to the power supply can then perform the transition to
$V^\prime$ within a safe slew rate, after which the tuning process is complete
and the application can start. This process is constrained within maximum and
minimum supported voltages and frequencies, based upon the power supply
capabilities, recommended limits and the operating ranges that were used in
order to generate safe values for $S_f$ and $S_v$.

\subsection{Characterisation}
\label{sec:characterisation}

Prior to testing this implementation, the RO characteristics $S_f$ and $S_v$ are
determined through empirical measurement of an XMOS XS1-U8A-64 processor, shown
in Table~\ref{tab:characterisation}.

The RO frequency in relation to voltage, $S_v$, is recorded, along with a
conservative $S_f$, the result of a series of frequency and voltage tests in
which a stress-test application was run to verify correct operation of the
hardware at each frequency/voltage combination. The stress-test exercises
multiple components of the processor simultaneously. It has three possible
outcomes: success, where the test completes without error; failure, where an
error is detected during execution; or crash, where the system becomes
unresponsive. The test is not considered a certification of reliability (it
comes with no guarantee from the vendor), but is sufficient for experimental
purposes; if this test does not exhibit transient faults, none are seen in
regular applications on the same test bed.

\begin{table}[h!]
\centering
\begin{tabular}{|c|c|}
\hline 
\textbf{Term} & \textbf{Value}
\\ \hline
$S_f$ & 1.7
\\ \hline
$S_v$ & $5.95\text{e}-06$
\\ \hline
\end{tabular}
\caption{The RO scaling terms for a XS1-U8A-64 processor.}
\label{tab:characterisation}
\end{table}

\section{Testing}
\label{sec:test}

Following the implementation of the voltage tuning framework, it was tested on
nine XS1-U8A-64 processors, covering three each of slow, typical and fast
silicon samples. A test rig, capable of reading the power supply voltages, was
used in order to accurately observe the changes in voltage that were applied.
For each test run, the tuning process was performed and the target voltage set
as per the description in Section~\ref{sec:imp}.

To verify system stability at the tuned voltage, a stress-test application is
used, as described in Section~\ref{sec:characterisation}. Prior to a full suite
of tests, the $S_f$ and $S_v$ parameters were tested on slow silicon to confirm
the parameters were chosen correctly to avoid failures or crashes on all samples
of the chip. The tuning framework and stress-test is run three times on each of
the nine processors and average values collected. It is worth noting, however,
that there was negligible variation between test runs on any given sample
processor. Tests were conducted at 500~MHz and 400~MHz to demonstrate the
capability to tune depending on the required system performance.

\begin{figure}
\centerline{\includegraphics[trim=0cm 0cm 0cm
0cm,clip=true,width=3.5in]{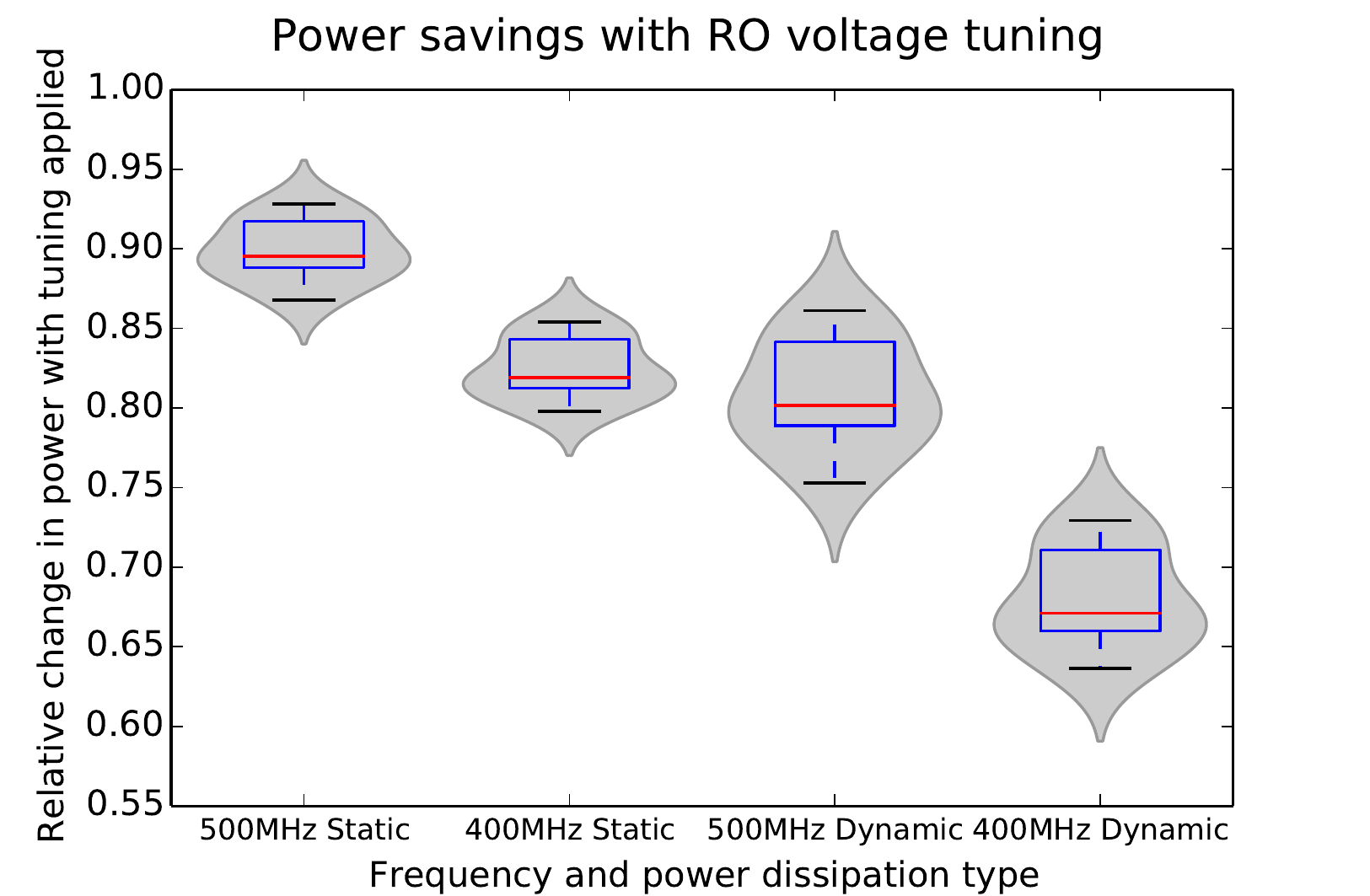}}
\caption{Violin plot of static and dynamic power savings for RO voltage tuning
at 500 and 400~MHz across a sample of nine XS1-U8A-64 chips. The top whiskers
represent the saving in the slowest silicon under test and the bottom whiskers
represent the fastest sample.}
\label{fig:savings}
\end{figure}

Figure~\ref{fig:savings} shows a box-plot of the reduction in static and dynamic
processor power for each of the sampled chips at 500~MHz and 400~MHz, compared
to the nominal operating voltage of 1 Volt. There is no voltage/frequency table
specific for this processor, so this is the typical operating point, per the
datasheet. The change in static and dynamic power is determined by evaluating
the voltage terms in Equations~\ref{eq:static} and~\ref{eq:dynamic}, whilst the
other terms remain unchanged. The figure also shows the kernel density plot of
the data beneath the box-plots, forming a violin plot and projecting the
behaviour for a larger sample set.

At 500~MHz, the slowest processor of the sample set benefits from a 70~mV
reduction in the core supply, yielding a dynamic power reduction to 0.86 of the
default and a static power reduction to 0.92 of its original state. The
fastest processor gets a 140~mV reduction, lowering dynamic power and static
power to 0.75 and 0.87 of their prior levels, respectively.

\begin{figure}
\centerline{\includegraphics[trim=0cm 0cm 0cm
0cm,clip=true,width=3.5in]{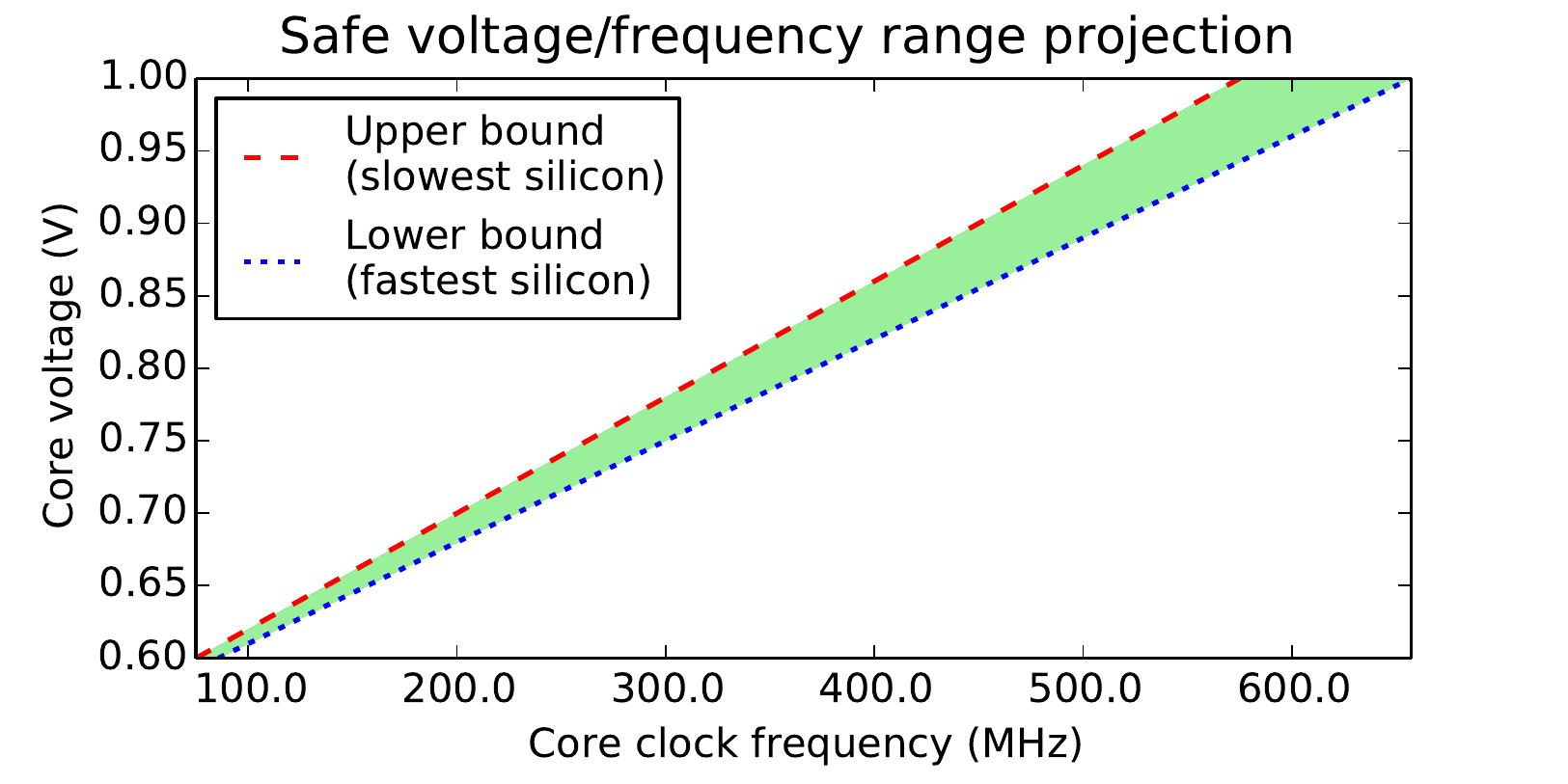}}
\caption{Projection of safe voltage/frequency combinations within 0.6--1.0~V
for the fastest and slowest silicon samples tested.}
\label{fig:projection}
\end{figure}

At 400~MHz, the power savings are greater and the distributions more spread out,
but follows the same shape as for 500~MHz, in line with the characteristics of
the sampled chips. Figure~\ref{fig:projection} projects the $F_p^\prime$ and
$V^\prime$ combinations across a wider range, demonstrating the range voltages
that would be applied to different chip samples for a particular $F_p^\prime$.

This data demonstrates that RO tuning can save system energy in all cases, but
most importantly, can save more energy in processors where the silicon is fast
enough to allow it. The achieved dynamic power saving across the sampled chips
varies by 14\% at the maximum operating frequency of 500~MHz, passing the
critical path test application in all cases.

\section{Evaluation}
\label{sec:ev}

Our RO based voltage tuning has been shown to be effective at reducing device
energy consumption with zero impact on program performance, save for a slower
start-up time. The method is able to save over 25\% of power in the core supply
for suitable samples of silicon and achieve power saving in all samples. The
actual power saving will depend on the behaviour of the application, but in a
typical scenario this may reduce the power dissipation from approximately 150~mW
to 110~mW, when considering the power profile of the XS1-U8A-64.

A conservative and linear relationship between RO frequency, target core
frequency and lowest safe voltage is used. As such, the energy saving at a given
frequency is not the absolute minimum, nor does the energy saving exactly fit
the curve of the silicon's performance as voltage and temperature change.
Specifically targeted hardware solutions are therefore able to better
characterise the critical path and provide tighter voltage tuning.  However,
this approach is still able to provide reliable operation whilst saving energy,
provided there are no severe variations in the operating conditions.

The strength of this approach is in its use of general purpose hardware for
characterisation and software for control. This creates a highly flexible
voltage tuning implementation that can be easily mapped to other similarly
equipped devices, doesn't interfere with the real-time behaviour of the running
application, and is unconstrained in voltage/frequency selection except for any
limitations imposed by the hardware.

\section{Future work}
\label{sec:future}

Areas of future work include more sophisticated software implementations for the
control loop, integration with different hardware critical path estimation
methods, and the application of this work to other architectures. This section
discusses all of these areas in turn.

\subsubsection*{Software DVFS}

The current software implementation is run before program startup, minimising
integration effort and guaranteeing no disruption to program execution. There is
scope for applying this technique in a periodic manner, continuing to sample RO
speed throughout program execution in order to adapt to environmental changes,
or changes in the workload of the processor that might create more or less heat.
On the current target architecture, this could be implemented as a dedicated
thread, provided the instrumented application's resource requirements (with
respect to number of threads and performance) would not be adversely affected.
This would allow our approach to be used as a continuous controller, in a
comparable manner to full hardware implementations such as~\cite{burd2000scaled,
das2006dvstuning, Liu2010}.

In some applications, it may be beneficial to apply the constraints in reverse.
For example, in an energy-scarce environment, a maximum voltage may be
available, and so the control loop should tune frequency appropriately,
maximising performance with the available voltage supply. This is a relatively
straightforward task in terms of engineering the control framework, although the
impact upon the performance of relevant applications would need to be studied.

\subsubsection*{Critical path characterisation}

One of the key contributions of this paper is the use of a software controlled
\emph{multi-purpose hardware block}, rather than a dedicated hardware block
designed for critical path representation. However, the software control loop
could be integrated with an appropriately instrumented critical path
representation such as the UDL~\cite{Ikenaga2011}, creating a more accurate
control loop that is still software driven.

In addition, techniques similar to monitor timing slack such as that
in~\cite{Hsieh2011} could be applied, although sufficiently fine grained and
accurate voltage samples may be impractical in a hardware-software control loop.

\subsubsection*{Other devices, direct comparison}

Wider comparisons could be drawn by applying this technique to a range of
architectures, starting by identifying those with similarly controllable RO
hardware. Of particular interest is FPGAs, with which work such
as~\cite{Nabina2012,Nunez-Yanez2013} could be directly compared to our method,
by instrumenting a design with each of the forms of sensing and control.

\section{Conclusions}
\label{sec:conclusions}

This paper has presented a technique for using multi-purpose ring oscillators to
provide a characterisation of device speed, accounting for process variation
across samples of a device and environmental factors such as temperature. This
characterisation is utilised by a software control loop, which tunes the
operating voltage of the device to a safe minimum at the required core clock
frequency. The result is a variation sensitive power optimisation method that
uses a simple hardware block and flexible software controller that can provide
significant power savings.

The characterisation and control method is demonstrated on a device with
software accessible ring oscillators, the XMOS XS1-U8A-64 embedded
microprocessor. In testing on a sample set of this processor, the method saves
between 14\% and 25\% of dynamic power, demonstrating sensitivity to silicon
speed and saving power in all test cases. 

The control method has zero impact on the performance of any application run
on the processor as the optimisation is performed at start-up. The software
implementation of this control method is available upon request to the
authors. Further work has been proposed that would allow testing of this
technique with other devices to allow closer comparison with other voltage
tuning methods. In addition, the flexibility of the software control
implementation could be leveraged to provide similar voltage tuning
optimisations using other types of characterisation hardware or as a
continuously operating control loop.

\section*{Acknowledgments}

The authors would like to thank Henk Muller and Jon Ferguson at XMOS for the
supply of hardware enabling this research.

\bibliographystyle{alphaurl}
\bibliography{../../../../bibs/EACO}

\end{document}